\documentclass[12pt]{article}
\usepackage{amsmath, latexsym, amssymb}

\newcommand{\R}{\mathbb R}
\newcommand{\C}{\mathbb C}

\newcommand{\p}{\partial}

\newcommand{\Xa}{\mathbf{X}_{(\alpha)}}

\newcommand{\PM}{\mathfrak{P}_{\mathrm{m}_0}^{\mathbb{V}}\mathbb{M}}

\newcommand{\DQ}[1]{\mathcal{D}_{{Q},{W}}#1}

\newcommand{\Bold}[1]{\mbox{\boldmath$\mathit{#1}$}}

\newcommand{\zdual}{\langle \mathfrak{b}',\mathfrak{b}\rangle}

\newcommand{\ti}{\mathrm{t}}
\newcommand{\ii}{\mathrm{i}}
\newcommand{\x}{\mathrm{x}}
\newcommand{\s}{\mathrm{s}}

\newcommand{\pp}{\mathfrak{p}}
\newcommand{\f}{\mathfrak{f}}

\newcommand{\m}{\mathrm{m}}
\newcommand{\dd}{\mathrm{d}}
\newcommand{\M}{\mathbb{M}}
\newcommand{\U}{\mathbb{U}}
\newcommand{\I}{\mathbb{I}}
\newcommand{\D}{\mathbb{D}}
\newcommand{\FF}{\mathbb{F}}
\newcommand{\PP}{\mathbb{P}}
\newcommand{\V}{\mathbb{V}}
\newcommand{\z}{\mathfrak{b}}
\newcommand{\Z}{\mathfrak{B}}
\newcommand{\F}{\mathfrak{F}}
\newcommand{\ttt}{\mathfrak{t}}
\newcommand{\TT}{\mathfrak{T}}
\newcommand{\QED}{\hspace{.2in}\square\newline}

\newtheorem{corollary}{Corollary}[section]
\newtheorem{proposition}{Proposition}[section]
\newtheorem{definition}{Definition}[section]
\newtheorem{assumption}{Restriction}[section]

\numberwithin{equation}{section}
\begin{document}

\title{Functional Integration for Quantum Field Theory}

\author{J. LaChapelle }

\maketitle
\begin{abstract}
The functional integration scheme for path integrals advanced by Cartier
and DeWitt-Morette is extended to the case of fields. The extended
scheme is then applied to quantum field theory. Several aspects of the
construction are discussed.
\end{abstract}
\noindent\sloppy{{\bf Keywords}: functional integration, quantum field
theory}


\section{Introduction}
Functional integration has proven its usefulness in both physics and
mathematics. It is expected that this usefulness will be considerably
extended if the subject can be rigorously formulated. Cartier and
DeWitt-Morette have advanced a scheme for this formulation
(\cite{CA/D-M},\cite{CA/D-W2},\cite{CA/D-M3}). However, for physical
applications, the scheme is restricted to functional integrals over
spaces of paths, i.e. maps defined on one-dimensional domains.

This paper proposes an extension of Cartier/DeWitt-Morette's scheme to
the case of functional integrals over spaces of fields, i.e. maps with
$d$-dimensional domains. It is a virtue of their framework that the
proposed alterations are minimal and fairly obvious.

The scope of the proposal is limited to a general scheme for fields of
even Grassmann parity and its correspondence with quantum field theory.
The study of a symplectic geometrical setting, fermionic fields,
renormalization, topological issues, and external/internal symmetry
issues will be developed and reported later.  It should be kept in mind
that the extension is a proposal: its correctness/usefulness can only be
established by extensive study and application. It is quite possible
that the details of the proposal will require modification.

\section{Functional Integration}\label{sec. functional integration}
In this section we briefly review the Cartier/DeWitt-Morette (CDM)
scheme for functional integration over spaces of pointed \emph{paths};
i.e. maps $\mathfrak{p}:(\mathbb{T},\ti_0)\rightarrow (\M,\m_0)$ where
$\mathbb{T}\subseteq\R$ and $\M$ is a differentiable manifold. We then
propose a generalization of their formulation to include spaces of
\emph{fields} $\mathfrak{f}:\D\rightarrow \M$ where $\D\subseteq\R^r$.

\subsection{CDM scheme}
The CDM scheme (see, for example, \cite{CA/D-M}, \cite{CA/D-W2}) defines
functional integrals given the data $(\Z,\Theta,Z,\mathcal{F}(\Z))$.
\subsubsection{Definitions}
Here $\Z$ is a separable (usually) infinite dimensional Banach space
with a norm $\|\z\|$ where $\z\in \Z$ is an $L^{2,1}$ map
$\z:[\ti_a,\ti_b]\in\R\rightarrow \M$ with $\M$ an $m$-dimensional
paracompact differentiable manifold. The dual Banach space $\Z'\ni \z'$
is a space of linear forms such that $\zdual_\Z\in\C$ with an induced
norm given by
\begin{equation}
  \|\z'\|=\sup_{\z\neq 0}|\zdual|/\|\z\|\;.\notag
\end{equation}
Assume $\Z'$ is separable. Then $\Z'$ is Polish and consequently admits
complex Borel measures $\mu$. $\Theta$ and $Z$ are bounded,
$\mu$-integrable functionals $\Theta:\Z\times \Z'\rightarrow \C$ and
$Z:\Z'\rightarrow\C$.

These data are used to define an integrator $\mathcal{D}_{\Theta,Z}\z$
on $\Z$ by
\begin{definition}\label{def1}
\begin{equation}\label{integrator}
  \int_{\Z}\Theta(\z,\z';\mathrm{s})\,\mathcal{D}_{\Theta,Z}\z:=Z(\z';\mathrm{s})
\end{equation}
\end{definition}
where $\mathrm{s}\in\C_+$ is a parameter. (Allowing the $\mathrm{s}$
parameter to reside in the complex half-plane is a slight extension of
the CDM scheme: They put $\mathrm{s}\in\{1,i\}$.)

\begin{quote}
\emph{Example}: The well-known Gaussian integrator is characterized by
the choices
\begin{eqnarray}
  &&\Theta(\z,\z';s)=\exp{\{-(\pi/s) Q(\z)-2\pi i{\langle
  \z',\z\rangle}_{\Z}\}}\notag\\
  &&Z(\z';s)=\exp{\{-\pi sW(\z')\}}
\end{eqnarray}
where $Q$ is a nondegenerate bilinear form on $\Z$ such that
$\mathrm{Re}\,(Q/s)>0$, and $W$ is its inverse on the dual space $\Z'$.
 A one parameter family of Gaussian integrators $\mathcal{D}\omega_s$
 is defined according to (\ref{def1}) by
\begin{equation}\label{gaussian}
  \int_{\Z}\exp^{-2\pi i{\langle
  \z',\z\rangle}}\mathcal{D}\omega_s(\z):=\int_{\Z}\exp^{\{-(\pi/s) Q(\z)-2\pi i{\langle
  \z',\z\rangle}\}}\,\DQ \z
  :=e^{-\pi sW(\z')}\;.
\end{equation}
This can be interpreted as a characterization of the Gaussian integrator
by its Fourier-Stieltjes transform;
\begin{equation}
\mathcal{F}\omega_s(\z'):=\exp^{-\pi sW(\z')}\;.
\end{equation}

\end{quote}

The final datum is the space of integrable functionals $\mathcal{F}(\Z)$
consisting of functionals\footnote{Assume that $F(\z;\mathrm{s})$ is (at
least) $C^1$ with respect to $\s$.} $F(\z;\mathrm{s})$ defined relative
to $\mu$ by
\begin{definition}\label{def2}
\begin{equation}\label{integrable functions}
F_{\mu}(\z;\mathrm{s}):=\int_{\Z'}\Theta(\z,\z';\mathrm{s})\,d\mu(\z')\;.
\end{equation}
\end{definition}
If $\mu\mapsto F_{\mu}$ is injective, then $\mathcal{F}(\Z)$ is a Banach
space endowed with a norm $\|F_{\mu}\|$ defined to be the total
variation of $\mu$ for all $\mathrm{s}$.
\begin{quote}
\emph{Example}: For the Gaussian case, note that
\begin{eqnarray}
F_{\mu}(\z;\mathrm{s}):=\exp^{-(\pi/s)Q(\z)}\int_{\Z'}\exp^{\{-2\pi
i{\langle \z',\z\rangle}\}}\,d\mu(\z')\notag\\
=\mathcal{F}\mu(\z)\exp^{-(\pi/s) Q(\z)}
\end{eqnarray}
where $\mathcal{F}\mu(\z)$ is the Fourier-Stieltjes transform of the
measure $\mu$ on $\Z'$.
\end{quote}

An integral operator $\int_\Z$ on the normed Banach space
$\mathcal{F}(\Z)$ is then defined by
\begin{definition}\label{def3}
\begin{equation}\label{integral definition}
  \int_{\Z}F_{\mu}(\z;\mathrm{s})\,\mathcal{D}_{\Theta,Z}\z:=
  \int_{\Z'}Z(\z';\mathrm{s})\,d\mu(\z')\;.
\end{equation}
\end{definition}
The integral operator $\int_\Z$ is a bounded linear form on
$\mathcal{F}(\Z)$ with
\begin{equation}
\left|\int_\Z
F_{\mu}(\z;\mathrm{s})\,\mathcal{D}_{\Theta,Z}\z\right|\leq\|F_{\mu}\|
\end{equation}
for all $\mathrm{s}$ (\cite{CA/D-W2}).

\begin{quote}
\emph{Example}: Continuing with the Gaussian case, this definition can
be written as
\begin{equation}\label{symmetric}
\int_{\Z}\mathcal{F}\mu(\z)\mathcal{D}\omega_s(\z)
:=\int_{\Z'}\mathcal{F}\omega_s(\z')\,d\mu(\z')\;.
\end{equation}
Moreover, the definitions could have been stated with
$\mathcal{D}_{\Theta,Z}\z\rightarrow\mathcal{D}\omega_s(\z)$. Then the
choice
\begin{eqnarray}
  &&\Theta(\z,\z')=\exp^{-2\pi i{\langle
  \z',\z\rangle}_{\Z}}\notag\\
  &&Z(\z';s)=\exp^{-\pi sW(\z')}
\end{eqnarray}
would again characterize the Gaussian integrator. But the integrable
functionals would be of the form $F_{\mu}(\z)=\mathcal{F}\mu(\z)$.
\end{quote}

Note that $\mu$ is not a component of the specified data. Hence, the
right-hand side of definition (\ref{integral definition}) is not a
prescription for calculating the left-hand side. Instead, using existing
functional integral `technology', one invariably makes use of some form
of localization to reduce the left-hand side to a finite dimensional
integral. Thus a `good' characterization encoded by the definitions must
reduce to the correct finite integral in any dimension. (This is not to
say that including $\mu$ in the data and then studying the right-hand
side of (\ref{integral definition}) directly would not be useful.)

\subsubsection{Restrictions}\label{restrictions}
With this in mind, it is useful to impose two restrictions on the
general framework outlined in the previous subsection in order to
implement the analogues of invariant measures and integration by parts
in finite dimensions. This suggests interpreting the previous
definitions in the context of differential forms. That is, interpret
$F_{\mu}(\z;\mathrm{s})\,\mathcal{D}_{\Theta,Z}\z=:F_{\mu}(\mathrm{s})$
as a volume element on $\Z$ and let $\mathcal{F}^{\wedge}(\Z)$ denote
the space of integrable volume elements on $\Z$.

Let $\mathfrak{Y}$ be a separable Banach space and
$M:\mathfrak{X}\rightarrow \mathfrak{Y}$ be a diffeomorphism with
derivative mapping
$M'_{(\mathfrak{x})}:T_\mathfrak{x}\mathfrak{X}\rightarrow
T_\mathfrak{y}\mathfrak{Y}$. The derivative map possesses a transpose
$\widetilde{M'}:T\mathfrak{Y}'\rightarrow T\mathfrak{X}'$ defined by
\begin{equation}
  \langle \widetilde{M'}_{(\mathfrak{y}')}\Bold{\mathfrak{y}}',\Bold{\mathfrak{x}}
  \rangle_{T_{\mathfrak{x}}\mathfrak{X}}
  :=\langle
  \Bold{\mathfrak{y}}',M'_{(\mathfrak{x})}\Bold{\mathfrak{x}}
  \rangle_{T_\mathfrak{y}\mathfrak{Y}}\;.
\end{equation}
$M$ induces a germ $R:\mathfrak{Y}'\rightarrow \mathfrak{X}'$ along with
its associated derivative mapping
$R'_{(\mathfrak{y}')}:T_{\mathfrak{y}'}\mathfrak{Y}'\rightarrow
T_{\mathfrak{x}'}\mathfrak{X}'$ and transpose
$\widetilde{R'}_{(\mathfrak{x})}:T_\mathfrak{x}\mathfrak{X}\rightarrow
T_\mathfrak{y}\mathfrak{Y}$ from the relation
\begin{equation}
  \langle R'_{(\mathfrak{y}')}\Bold{\mathfrak{y}}',
  \Bold{\mathfrak{x}}\rangle_{T_\mathfrak{x}\mathfrak{X}}=\langle
  \Bold{\mathfrak{y}}',\widetilde{R}'_{(\mathfrak{x})}
  \Bold{\mathfrak{x}}\rangle_{T_\mathfrak{y}\mathfrak{Y}}\;.
\end{equation}

Define the $\mu$-integrable functionals
$\overline{\Theta}:\mathfrak{Y}\times \mathfrak{Y}'\rightarrow\C$ and
$\overline{Z}:\mathfrak{Y}'\rightarrow \C$ by
\begin{equation}\label{theta bar}
  \overline{\Theta}\circ (M\times R^{-1}):=\Theta
\end{equation}
and
\begin{equation}\label{Z bar}
  \overline{Z}\circ R^{-1}:=Z
\end{equation}
along with their associated integrator
\begin{equation}
  \int_{\mathfrak{Y}}\overline{\Theta}(\mathfrak{y},\mathfrak{y}';\mathrm{s})
  \,\mathcal{D}_{\overline{\Theta},\overline{Z}}\mathfrak{y}
  :=\overline{Z}(\mathfrak{y}';\mathrm{s})\;.
\end{equation}

If $R(\mathfrak{Y}')=\mathfrak{Y}'$, enforce the transformation property
consistent with the volume element interpretation by requiring
\begin{assumption}\label{theta relation}
\begin{equation}\label{theta relation eq.}
  \overline{Z}(\mathfrak{y}';\mathrm{s})
  =\left(\mathrm{Det}R'_{(\mathfrak{y}')}\right)^{-1}Z(\mathfrak{y}';\mathrm{s})
\end{equation}
\end{assumption}
for $R$ a diffeomorphism and non-vanishing
$\mathrm{Det}R'_{(\mathfrak{y}')}$.\footnote{Note that this does not
include the absolute value as proposed in \cite{LA2}.} The determinant
is assumed to be well-defined and properly regularized.

Again, to be consistent with the volume element interpretation and to
indirectly implement integration by parts, require
\begin{assumption}\label{top form}
\begin{equation}\label{closed form eq.}
  d[\mathcal{D}_{\Theta,Z}\z\,]=0\;.
\end{equation}
\end{assumption}

Besides helping to characterize invariant integrators and integration by
parts, these two restrictions characterize a distinguished class of
integrable volume elements. Denote by $\mathcal{F}_R^{\wedge}(\Z)$ the
class of integrable volume elements that satisfy
\begin{equation}\label{integral invariant}
\int_{\Z}\mathcal{L}_{\Bold{\mathfrak{v}}}F_{\mu}(\mathrm{s})
=0\;\;\forall \mathrm{s}
\end{equation}
where $\Bold{\mathfrak{v}}$ is the generator of a group action
$\sigma_g:\Z\rightarrow \Z$ such that
$\langle\sigma_g'\z',\z\rangle=\langle \z',\sigma_g\z\rangle$. Likewise
denote by $\mathcal{F}_R(\Z)$ the class of integrable functionals
$F_{\mu}(\z;\mathrm{s})$ associated with $\mathcal{F}_R^{\wedge}(\Z)$
through the integrator $\mathcal{D}_{\Theta,Z}\z$. Because of
Definitions \ref{def1}--\ref{def3}, this translates to invariance
conditions on $\Theta(\z,\z')$ and $Z(\z')$.

In many applications of interest, it turns out that the integrators of
interest are invariant under some group action on $\Z$, i.e.,
$\mathcal{L}_{\Bold{\mathfrak{v}}}[\mathcal{D}_{\Theta,Z}\z\,]=0$, i.e.,
$\mathcal{D}_{\Theta,Z}\z$ is an absolute integral invariant of the
group action. This puts
 constraints on the integrable functionals $F_{\mu}(\z;\mathrm{s})$. Note that
(\ref{integral invariant}) may be quite restrictive and
$\mathcal{F}_R^{\wedge}(\Z)$ may be severely limited or even empty.

\begin{quote}
\emph{Example}: For example, let $\sigma_g\z=\z+\z_0$ with $\z_0$ a
fixed but arbitrary point in $\Z$ and $\sigma'_g\z'=\z'-\z'_0$ where
$\langle \z_0',\z\rangle:=-\langle \z',\z_0\rangle$. From Definition
\ref{def3} and the fact that Restriction \ref{theta relation} implies
$\overline{Z}(\z'-\z_0')=Z(\z'-\z_0')$, it follows that
$\mathcal{D}_{\Theta,Z}\z$ is an absolute integral invariant of
$\sigma_g$. Hence, if
$F_{\mu}(\mathrm{s})\in\mathcal{F}_R^{\wedge}(\Z)$, then
\begin{equation}
\int_{\Z}\mathcal{L}_{\Bold{\mathfrak{v}}}F_{\mu}(\mathrm{s})
=\int_{\Z}\mathcal{L}_{\Bold{\mathfrak{v}}} [F_{\mu}(\z;\mathrm{s})]
\mathcal{D}_{\Theta,Z}\z= \int_{\Z}\mathrm{i}_{\boldmath{\mathfrak{v}}}
d[F_{\mu}(\z;\mathrm{s})] \mathcal{D}_{\Theta,Z}\z=0\;.
\end{equation}

Since $\Bold{\mathfrak{v}}=\z_0$ is arbitrary, we have using Proposition
\ref{prop. integration by parts},
\begin{equation}
\int_{\Z}\frac{\delta F_{\mu}(\z;\mathrm{s})}{\delta
\z^i(\dd)}\mathcal{D}_{\Theta,Z}\z =-\int_{\Z}
F_{\mu}(\z;\mathrm{s})\frac{\delta}{\delta
\z^i(\dd)}\mathcal{D}_{\Theta,Z}\z =0\;.
\end{equation}
\end{quote}

\subsubsection{Parametrization}
Up to this point, the construction relies on the fact that $\Z$ is a
Banach space. However, for many interesting applications, the goal is to
define an integral over an infinite dimensional space,
$\mathfrak{P}_{\m_0}(\mathbb{M})$, composed of $L^{2,1}$ pointed
paths\footnote{Pointed paths are not the only cases of interest:
$\pp:S^1\rightarrow \M$ and $\pp:\R\rightarrow \M$ are important cases.}
$\pp:(\mathbb{T},\ti_0)\rightarrow (\M,\m_0)$ where
$\mathbb{T}=[\ti_a,\ti_b]$, $\M$ is an $m$-dimensional differentiable
manifold, and $\m_0$ is some fiducial point. Unfortunately,
$\mathfrak{P}_{\m_0}(\mathbb{M})$ is not generally a Banach space.

To remedy this, CDM introduce a set of $n\leq m$ linearly independent
vector fields $\Xa$ where $\alpha\in\{1,\ldots,n\}$. The set $\{\Xa\}$
generates a sub-bundle (distribution) $\V\subseteq T\M$. Then the space
$\PM$ can be parametrized by $\mathfrak{P}_0(\C^n)$ via the differential
system
\begin{equation}\label{parametrization}
  \left\{ \begin{array}{ll}
    \Bold{d\pp}(\ti,\z)-\mathbf{Y}(\pp(\ti,\z))\Bold{d\ti}=
    {\Xa}(\pp(\ti,\z))\Bold{d\z}(\ti)^{\alpha} \\
    \pp(\ti_0)=\m_0
  \end{array}
  \right.
\end{equation}
where $\z\in\mathfrak{P}_0(\C^n):(\mathbb{T},\ti_0)\rightarrow(\C^n,0)$,
$\dot{\Bold{\pp}}(\ti)-\mathbf{Y}(\pp(\ti,\z))\in \V_{\pp(\ti)}\subseteq
T_{\pp(\ti)}\M$ for each $\ti\in\mathbb{T}$, and $\mathbf{Y}$ is some
given vector field on $\M$. The solution of (\ref{parametrization}) is
denoted $\pp(\ti,\z)=\m_0\cdot \mathit{\Sigma}(\ti,\z)$. If the
transformation $\pp\mapsto\z$ can be inverted, the differential system
induces a parametrization $P:\mathfrak{P}_0(\C^n) \rightarrow \PM$.

Fortunately $\mathfrak{P}_0(\C^n)$ is a Banach space. Together with
Definition \ref{def3} this allows for a definition of the path integral
of a functional $F(\pp;\ti)$ over $\PM$:
\begin{definition}\label{def4}
\begin{equation}\label{PM integral def.}
  \int_{\PM}F(\pp;\ti)\, \mathcal{D}\pp :=
   \int_{\mathfrak{P}_0(\C^n)}F_{\mu}(\m_0\cdot \mathit{\Sigma}(\ti,\z))\,
  \mathcal{D}_{\Theta,Z}\z\;.
\end{equation}
\end{definition}

\subsection{Generalization to Fields}
\subsubsection{CDM Revisited}
It is useful to reformulate and reinterpret certain aspect of the CDM
scheme as it suggests a generalization to fields. The idea is to remove
from the definitions (as much as possible) any dependence on the source
and target manifolds of $\z$ and the parameter $\s$.

Step 1): Make an affine transformation on $\mathbb{T}$ so that
$\mathbb{T}\rightarrow[0,1]=:\I$. This induces a map on $\Z$ by
$\z\mapsto\z/\sqrt{(\ti_b-\ti_a)}=:\widetilde{\z}$ since $\z$ is
required to be $L^{2,1}$. From the duality
\begin{equation}
\zdual_\Z:=\int_{\mathbb{T}}\z'(\ti)\z(\ti)d\ti\;,
\end{equation}
deduce that $\z'\mapsto\sqrt{(\ti_b-\ti_a)}\z'=:\widetilde{\z}'$.

The parametrization in terms of
$\widetilde{\z}\in\mathfrak{P}_0(\C^n):(\I,\ii_0)\rightarrow(\C^n,0)$
becomes
\begin{equation}\label{scaled parametrization}
  \left\{ \begin{array}{ll}
    \Bold{d\pp}(\ii,\z)-(\ti_b-\ti_a)\mathbf{Y}(\pp(\ii,\z))\Bold{d\ii}=
    \sqrt{(\ti_b-\ti_a)}{\Xa}(\pp(\ii,\z))\Bold{d\widetilde{\z}}(\ii)^{\alpha} \\
    \pp(\ii_0)=\m_0
  \end{array}
  \right.
\end{equation}

Step 2): Choose the functionals $\Theta$ and $Z$ so that the factor
$(\ti_b-\ti_a)$ can be absorbed into the parameter $\s$.

\begin{quote}\emph{Example}: Returning to the Gaussian integrator,
observe that the parameter $\s$ is situated just right to absorb factors
of $(\ti_b-\ti_a)$. Notice that $\Theta(\z,\z';(\ti_b-\ti_a))
=\Theta(\widetilde{\z},\widetilde{\z'};1)$ and
$Z(\z';(\ti_b-\ti_a))=Z(\widetilde{\z'};1)$ follows from the scaling
properties of $\z$ and $\z'$.
\end{quote}

The first two steps transfer the dependence on $\mathbb{T}$ to a
dependence on $\s$, but this is still not satisfactory. However, it does
suggest the next step.

Step 3): Imagine that the $\s$ parameter is a result of a localization
in some function space. That is, augment the data by
$\Z\rightarrow\Z\otimes\TT$ where $\TT$ is a Banach space of pointed
maps $\ttt:(\I,\ii_0)\rightarrow(\C_+,0)$, and (somehow) determine new
$\Theta$ and $Z$ that will characterize a suitable integrator on
$\Z\otimes\TT$ (strictly, the completion of $\Z\otimes\TT$).

This step brings the parametrization to the form
\begin{equation}\label{augmented parametrization}
    \left\{ \begin{array}{ll}\Bold{d\pp}(\z,\ttt)(\ii)=
    {\Xa}(\pp(\z,\ttt))\Bold{d\z}(\ii)^{\alpha}
    +\mathbf{Y}(\pp(\z,\ttt))\Bold{d\ttt}(\ii)\\
    \pp(\z,\ttt)(\ii_0)=\m_0\;,
    \end{array}
    \right.
\end{equation}
and the functional integral definition becomes
\begin{equation}\label{new integral def.}
  \int_{\PM}F(\pp)\, \mathcal{D}\pp :=
   \int_{\mathfrak{P}_0(\C^n\times\C_+)}
   F_{\mu}(P(\z,\ttt))\,
  \mathcal{D}_{\Theta,Z}(\z,\ttt)\;.
\end{equation}

This effectively removes the source and target manifolds from the
intrinsic data $(\Z,\Theta, Z)$. Any remaining dependence resides in
$F_{\mu}$: Experience with standard applications suggests that the
functional $F_{\mu}$ will have to encode some kind of localization in
$\Z\otimes\TT$ to recover expected results. Conclude that
$\mathcal{F}(\Z\otimes\TT)$ (and hence $\Z'\otimes\TT'$ and/or $\mu$) is
where the source and target manifolds exert their influence.

\subsubsection{Field Formulation}
We are now in a position to propose a field generalization of the CDM
scheme.

The first generalization is simple; let $\Z$ be the Sobolev space
$W^{k,p}(\U)$ of maps $\z:\U\rightarrow\C^n$ with $\U\subseteq\R^d$ open
and unbounded or $\U\subseteq \mathbb{S}^d$. If $\U$ has boundary
$\p\U$, take $\Z=W_0^{k,p}(\U)$, the closure of the vector space of
$C^{\infty}$ maps with compact support in $\U$. Since $W^{k,p}(\U)$ and
$W_0^{k,p}(\U)$ are Banach, the first three definitions and the two
restrictions (with the $\s$ parameter removed) can be applied directly.
If $\Z_{(r)}=\bigotimes_i^r\Z_i$, then the measures on $\Z_{(r)}'$ are
defined to be the convolution of the corresponding measures on each
$\Z_i'$.

Let $\F(\M)$ denote the space of fields $\f:\D\rightarrow\M$ where $\D$
is a $d$-dimensional differentiable manifold. The second generalization
is to replace the parametrization (\ref{parametrization}) with the
exterior differential system
\begin{equation}
\left\{\Bold{\omega}_{I}=0\right\}
\end{equation}
where $\Bold{\omega}_{I}\in\Lambda T^*\F(\M)$ with $I\in\{1,\ldots,N\}$.
This system defines a parametrization $P:\Z\rightarrow\F(\M)$ by
\begin{equation}
P^*\Bold{\omega}_{I}=0\;\;\forall\; I\;.
\end{equation}
Two particularly prevalent parametrizations arise from Pfaff exterior
differential systems associated with the development map
$\mathrm{Dev}:\mathfrak{F}T_{\m_0}\M\rightarrow\F(\M)$ and the
exponential map $\mathrm{Exp}:T_{\f}\F(\M)\rightarrow\F(\M)$. (These are
just the functional counterparts to the well-known Cartan development
map and the exponential map.)

Finally, replace the definition \ref{def4} with
\begin{equation}\label{field integral def.}
  \int_{\F(\M)}F(\f)\, \mathcal{D}\f :=
   \int_{\Z}F_{\mu}(P(\z))\,
  \mathcal{D}_{\Theta,Z}(\z)\;.
\end{equation}
Of course, particular applications often require consideration of some
type of boundary conditions if $\p\D\neq\emptyset$, in which case
$\F(\M)$ and hence $\Z$ will need to be suitably restricted; but in
general we will not require pointed maps in $\Z$. Additionally,
particular applications may require restrictions on the rank and nature
of $P$ so as to obtain immersions, embeddings, regular embeddings, etc.
However, no restrictions are imposed in general, because one can
anticipate that interesting effects will be associated with non-constant
rank and irregular points of $P$.

\section{Application to Quantum Field Theory}
\subsection{Geometrical Setting}
Before exploring the application of the proposed field generalization to
quantum field theory, it is useful to formulate in detail the
geometrical setting. To maintain generality, we consider the case of
gauge field theory, outline the relevant constructions, and recall some
pertinent facts.

\begin{itemize}
\item
$(\M,\D,\pi,\FF,G)$: a fiber bundle $\pi:\M\rightarrow\D$ with typical
fiber $\FF$ and structure group $G$.
\begin{itemize}
\item
$\FF$ furnishes a realization $\rho$ of $G$.
\item
Given a local trivialization $(\U_i,\phi_i)$ of $\M$ where
$\phi_i(\m)=(\pi(\m),\hat{\phi}_i(\m))$ and a set of local sections
$\sigma_i:\U_i\rightarrow\pi^{-1}(\U_i)$, the local representative of a
(matter) field $\f$ on $\D$ (in the physics sense) can be characterized
by $\f_i(\mathrm{d}):=(\hat{\phi}_i\circ\sigma_i)(\mathrm{d})$.
\end{itemize}

\item
$(\PP,\D,\Pi,G)$: the principal bundle associated with
$(\M,\D,\pi,\FF,G)$.
\begin{itemize}
\item
associated with $\mathrm{p}\in \PP$ is an admissible map
$p:\FF\rightarrow\pi^{-1}(\mathrm{d})$, i.e. $\hat{\phi}_i\circ p=g$ for
some $g\in G$.
\item Let $\widetilde{\f}:\PP\rightarrow \FF$ denote an equivariant map,
i.e. $\widetilde{\f}(\mathrm{p}\cdot
g)=\rho(g^{-1})\widetilde{\f}(\mathrm{p})\;\forall \mathrm{p}\in
\PP,\;g\in G$.
\item
The space of fields can be identified with the space of equivariant maps
$\widetilde{\f}:\PP\rightarrow\FF$ since $\widetilde{\f}\circ\Pi^{-1}
=p^{-1}\circ\sigma_i\;\;\forall\mathrm{d}\in\U_i$.
\item
Given a local trivialization $(\U_i,\Phi_i)$ of $\PP$ where
$\Phi_i(\mathrm{p})=(\Pi(\mathrm{p}),\hat{\Phi}_i(\mathrm{p}))$, a
canonical section $s_i$ is defined by $\Phi_i\circ s_i=(\mathrm{p},e)$
where $e$ is the identity element in $G$. The local representative of a
field $\f$ on $\D$ can also be characterized by
$\widetilde{\f}(s_i(\mathrm{d}))=\widetilde{\f}(\mathrm{p})
=p^{-1}\circ\sigma_i(\mathrm{d})
=(\hat{\phi}_i\circ\sigma_i)(\mathrm{d}) =\f_i(\mathrm{d})$.
\item
Under a fiber automorphism
$Aut:\pi^{-1}(\mathrm{d})\rightarrow\pi^{-1}(\mathrm{d})$, equivariant
maps transform according to $Aut^*\widetilde{\f}(\mathrm{p})=
\widetilde{\f}(Aut(\mathrm{p}))
=\rho(g_{\mathrm{p}}^{-1})\widetilde{\f}(\mathrm{p})$ where
$g_{\mathrm{p}}$ is unique since $G$ acts freely in
$\pi^{-1}(\mathrm{d})$. \item If the principal bundle is equipped with a
connection $\Bold{\omega}$, $Aut$ induces the transformation
$A^*\Bold{\omega}(\Bold{v})
=\mathrm{Ad}(g_{\mathrm{p}}^{-1})\Bold{\omega}(\Bold{v})
+\Bold{\theta}_{\mathrm{MC}}(g'_{\mathrm{p}}\Bold{v})
\;\forall\Bold{v}\in T_{\mathrm{p}}\PP$.
\item
The connection pulls back to a gauge potential $\Bold{A}_i$ via the
canonical section $s_i$.
\end{itemize}

\item
$J^k(\M)$: the $k$-jet manifold of $\M$.
\begin{itemize}
\item
For open sets $\U_i\subset\D$ and a local sections $\sigma_i$,
$j^k_{\mathrm{d}}\sigma_i$ is the $k$-jet at $\mathrm{d}\in\U_i$ with
representative $\sigma_i$.
\item
$J^k(\M)
:=\{j^k_{\mathrm{d}}\sigma\;:\mathrm{d}\in\D,\;\sigma\in\Gamma(\D,\M)\}$.
\end{itemize}

\item
$(J^k(\M),\D,\pi^k,\mathrm{pr}^{(k)}G))$: the $k$-jet bundle of $\M$
where $\mathrm{pr}^{(k)}G$ is the $k$-th prolongation of the structure
group $G$.
\begin{itemize}
\item
$\pi_k:J^k(\M)\rightarrow\D$ by
$j^k_{\mathrm{d}}\sigma\mapsto\mathrm{d}$.
\item
a trivialization $(\U_i,\phi_i)$ of $(\M,\D,\pi,\FF,G)$ induces a
trivialization $(\U_i,j^k\phi_i)$ of $(J^k(\M),\pi_k,\D)$.
\item
For open sets $\U_i\subset\D$ and a local sections $\sigma_i$, the
sections $j^k\sigma_i:\D\rightarrow J^k(\M)$ are defined by
$j^k\sigma_i(\mathrm{d}):=j^k_{\mathrm{d}}\sigma_i$.
\item
$\mathrm{pr}^{(k)}\varphi:=(j^k\hat{\phi}\circ j^k\sigma)$ is the $k$-th
prolongation of $\varphi$.
\item
The $k$-th prolongation of $G$ is defined relative to a local
trivialization by
$\rho(\mathrm{pr}^{(k)}g^{-1})\mathrm{pr}^{(k)}\varphi_i(\mathrm{d})
:=\mathrm{pr}^{(k)}[\rho(g^{-1}){\varphi}_i(\mathrm{d})]\;\forall
\mathrm{d}\in \U_i,\;g\in G$
\end{itemize}

\item
$(J^k(\PP),\D,J^k\Pi,\mathrm{pr}^{(k)}G)$: the principal bundle
associated with $(J^k(\M),\pi_r,\D)$.
\begin{itemize}
\item
Let $j^k\widetilde{\varphi}$ be an equivariant map on $J^k(\PP)$. The
prolongation of $G$ can alternatively be  defined by
$j^k\widetilde{\varphi}(\mathrm{p}^k\cdot\mathrm{pr}^{(k)}g)
:=\mathrm{pr}^{(k)}\widetilde{\varphi}(\mathrm{p}^k\cdot
g)\;\forall\mathrm{p}^k\in J^k(\PP),\;g\in G$.
\end{itemize}

\item$\F(\M)$: the space of matter fields taking their values in $\M$.
\begin{itemize}
\item
The space of matter fields consists of all allowed sections
$\f:\D\rightarrow\M$, i.e. $\Gamma(\D,\M)$; or, equivalently, all
allowed equivariant maps $\widetilde{\f}:\PP\rightarrow\FF$.
\item
The tangent space at a particular field $T_{\f}\F(\M)$ is a Banach
space. Elements $\Bold{\mathfrak{v}}_{\f}\in T_{\f}\F(\M)$ are linear
maps $\Bold{\mathfrak{v}}_{\f}:T\D\rightarrow T\M$.
\item
An action functional on $\F(\M)$ is defined by
$\mathcal{S}(\mathfrak{f})=\int_{\D}(j^k\f)^*\mathcal{L}_{J^k(\M)}$
where $\mathcal{L}_{J^k(\M)}$ is a (Lagrangian) volume density on
$J^k(\M)$.
\end{itemize}
\item
$\mathfrak{G}(\mathfrak{g})$: the space of gauge potentials taking their
values in the Lie algebra $\mathfrak{g}$ of $G$.
\begin{itemize}
\item
The group $\mathcal{G}$ of gauge transformations act by right action on
$\mathfrak{G}(\mathfrak{g})$. Its Lie algebra is isomorphic to
$\mathfrak{g}\otimes C^{\infty}(\D,\R)$.
\item
An action functional on $\mathfrak{G}(\mathfrak{g})$ is defined by
$\mathcal{S}(\mathfrak{a})=\int_{\D}(j^k\mathfrak{a})^*\mathcal{L}_{J^k(\PP)}$.
\end{itemize}

\end{itemize}

\subsection{Matter Fields}
The data $(\mathfrak{B},\Theta,Z,\mathcal{F}(\Z))$ for matter fields
must be determined. Given source and target manifolds $\D$ and $\M$,
construct $(\M,\D,\pi,G)$. Equip $\M\stackrel{\pi}{\rightarrow}\D$ with
a nondegenerate quadratic form $\widetilde{q}$ with positive real part.
This induces a quadratic form $\widetilde{Q}$ on $\F(\M)$ by
\begin{equation}
\widetilde{Q}(\f):=\int_{\D}\widetilde{q}(\f(\dd))d\tau(\dd)
\end{equation}
where $d\tau$ is a relevant volume element on $\D$. Invariably this
$\widetilde{Q}$ is derived from the action functional $\mathcal{S}$.

Select an open set $\U\subseteq\C^n$.  For simplicity and because this
covers the majority of physics applications, restrict to
$\Z=W^{1,2}(\U)\equiv H^1(\U)$ and $\Z'=H^{-1}(\U)$ its (continuous)
dual. If $\p\D\neq\emptyset$, then take $\Z=W_0^{1,2}(\U)\equiv
H^1_0(\U)$. A distinct $\Z$ must be introduced for each
\emph{independent} real field component degree of freedom. Hence, in
general the total space will have the form
$\Z_{(r)}=\bigotimes_i^r\Z_i$. For example, the associated Banach space
for a complex scalar field has the form $\Z=\Z_1\otimes\Z_2$.

Construct an exterior differential system
\begin{equation}
\left\{\Bold{\omega}_{I}=0\right\}
\end{equation}
 where
$\Bold{\omega}_{I}\in\Lambda T^*\F(\M)$ with $I\in\{1,\ldots,N\}$.
Thereby induce a parametrization $P:\Z\rightarrow\F(\M)$ by
\begin{equation}
P^*\Bold{\omega}_{I}=0\;\;\forall\; I\;.
\end{equation}
Use $P$ to equip $\Z$ with a nondegenerate quadratic form $Q(\z)$ with
$\mathrm{Re}(Q)>0$;
\begin{equation}
Q(\z):=P^*\widetilde{Q}(\z)=\widetilde{Q}(P(\z))\;.
\end{equation}
By the Riesz theorem, there exists an (anti)isomorphism
$D:\Z\rightarrow\Z'$ such that $Q(\z)=\langle D\z,\z\rangle$.
Conversely, there exists a $G$ such that $DG=Id$, and $G$ defines a
quadratic form on $\Z'$ by $W(\z'):=\langle\z',G\z'\rangle$.

The physical interpretation of $\int_{\Z}\mathcal{D}\omega(\z)$
presented in the appendix, and the identification of a useful basis in
$\Z$ determine, to some extent, the choice of $\Theta$ and $Z$. For
field theory applications, it is not difficult to anticipate that a
useful basis for matter fields will be given by eigenvectors of creation
and annihilation operators.

With this as motivation, we introduce the Hermite integrator. Let
\begin{eqnarray}
  \Theta_n(\z,\z')&:=&\left(\mathrm{Det}\,\frac{\pi
W}{2}\right)^{n/2}
    H_n\left(\sqrt{\pi Q(\z)}\right)\exp^{\{-\pi
  Q(\z)-2\pi i\zdual\}}\nonumber\\
  &=:&
  \widehat{H}_n(\hat{\z})\exp^{\{-\hat{\z}^2-2\pi i\zdual\}}
\end{eqnarray}
with $H_n(\hat{\z})$ the $n$-th order \emph{functional} Hermite
polynomial. For $\Z_{(r)}:=\bigotimes_i^r\Z_i$, the convolution of
measures on $\Z'_{(r)}$ leads to
\begin{equation}
\widehat{H}_n(\hat{\z}_1,\hat{\z}_2,\ldots,\hat{\z}_r)
=\widehat{H}_n(\hat{\z}_1)\widehat{H}_n(\hat{\z}_2)\times\cdots\times
\widehat{H}_n(\hat{\z}_r)
\end{equation}

Let
\begin{equation}
  Z_n(\z'):=
  \int_{\Z'}\widehat{H}_{n}'\left(\sqrt{\pi s W(\widetilde{\z}')}\right)
  \exp^{-\pi s W(\z'-\widetilde{\z}')}
  \;d\mu (\widetilde{\z}')
\end{equation}
where
\begin{equation}
  \left.\widehat{H}_{n}'(\z')
  :=\widehat{H}_n(\z)\right|_{\z^j\rightarrow\delta^{(j)}(\z')}\;.
\end{equation}

The $n$-th order Hermite integrator is characterized by
\begin{equation}
\int_{\Z}\exp^{\left\{-2\pi\imath\zdual\right\}}
\mathcal{D}\rho_n(\z):=\int_{\Z}\widehat{H}_n(\hat{\z})
\exp^{\left\{-\hat{\z}^2
-2\pi\imath\zdual\right\}}\DQ{\z}=Z_n(\z')\;.\notag\\
\end{equation}
It is normalized according to
\begin{equation}\label{Hermitian normalization}
  \int_{\Z}\,\mathcal{D}\rho_n(\z)
  =\left\{\begin{array}{cl}
  1&\mbox{for}\,n=0\\
  0&\mbox{otherwise}
  \end{array}\right.\;,
\end{equation}
and $\mathcal{D}\rho_0({\z})$ is the familiar Gaussian integrator. The
`vacuum', relative to boundary conditions implicit in the
parametrization $P$, is defined by
\begin{equation}
\langle0|0\rangle_P:=\int_{\Z}\,\mathcal{D}\rho_0(\z)
\end{equation}

Integrable functionals are of the form
\begin{eqnarray}
  F_{\mu}(\z;n)&=&\widehat{H}_n(\hat{\z})\exp^{-\hat{\z}^2}\int_{\Z'}
  \exp^{\left\{-2\pi\imath\zdual\right\}}\,d\mu({\z}')\notag\\
  &=&\mathcal{F}\mu(\z)\widehat{H}_n(\hat{\z})\exp^{-\hat{\z}^2}\notag\\
  &=:&f(\z)\widehat{H}_n(\hat{\z})\exp^{-\hat{\z}^2}\;.
\end{eqnarray}

Under the linear map $L:\Z\rightarrow\R$ by $\z\mapsto\zdual\in\R$, the
quadratic form becomes
\begin{equation}
  (Q\circ L^{-1})(\zdual)=:Q_{\R}(\zdual)
  =\zdual^2/W({\z}')\;.
\end{equation}
So that, by Proposition \ref{prop. change of variable},
\begin{eqnarray}\label{1-dim Hermite}
\int_{\Z}\,\mathcal{D}\rho_n(\z)
&\longrightarrow&\left(\frac{\pi}{2}\right)^{n/2}\left(W({\z}')\right)^{(n-1)/2}\nonumber\\
&&\hspace{.5in}\times
\int_{\R}H_n\left(\begin{array}{l}\sqrt{\frac{\pi}{W({\z}')}}\end{array}
\mathrm{u}\right)
\exp^{\left\{-\pi\mathrm{u}^2W({\z}')\right\}}\,d\mathrm{u}\;.
\end{eqnarray}
More generally, let $L:\Z\rightarrow\R^m$ by $\z\mapsto
\mathbf{u}=(\langle\z'_1,\z\rangle,\ldots,\langle\z'_m,\z\rangle)$ and
denote $W_{\R^m}=(W\circ\widetilde{L})$, then $Q_{\R^m}(\mathbf{u})=
[W_{\R^m}^{-1}]_{ij}\mathrm{u}^i
\mathrm{u}^j:=W_{\R^m}^{-1}(\mathbf{u})$ and
\begin{eqnarray}\label{n-dim Hermite}
  \int_{\Z}f(\z)\;\mathcal{D}\rho_n(\z)
&=&\left(\mathrm{det}\,\frac{\pi
W_{\R^m}}{2}\right)^{n/2}(\mathrm{det}\,W_{\R^m})^{-1/2}\nonumber\\
&&\hspace{.0in}\times\int_{\R^m}f(\mathbf{u}) H_n\left(\begin{array}{l}
\sqrt{\pi W_{\R^m}^{-1}(\mathbf{u})}\end{array} \right)
\exp^{\{-(\pi)W_{\R^m}^{-1}(\mathbf{u})\}}\;d\mathbf{u}\nonumber\\
\end{eqnarray}
where $H_n(\mathbf{u}):=\Pi_{i=1}^{m}H_{\alpha_{i}}(\mathrm{u}^i)$ with
$|\alpha|:=\Sigma_{i=1}^m(\alpha_i)=n$ is an $m$-fold product of Hermite
polynomials (see e.g. \cite{DU/XU}). Of particular interest are
functions of the form
\begin{equation}
f_m(\zdual):=\left(\frac{\pi W(\z')}{2}\right)^{m/2}H_m\left(\sqrt{\pi
Q(\zdual)}\right)
\end{equation}
for which
\begin{equation}
\int_{\Z}f_m(\zdual)\;\mathcal{D}\rho_n(\z)=\left(\pi
W(\z')\right)^mn!\,\delta_{nm}\;.
\end{equation} The relationship with Hermite
polynomials, and the interpretation of
$\int_{\Z}\,\mathcal{D}\rho_n(\z)$ as a linear form on $\mathcal{F}(\Z)$
is manifest.

In terms of half-densities associated with $\mathcal{D}\rho_n({\z})$, we
can make the formal correspondence
\begin{equation}
\varphi_n(\z)\equiv\widehat{H}_n(\hat{\z})
\sqrt{\mathcal{D}\rho_n({\z})}\;.
\end{equation}
According to the characterization, for $\z\mapsto\zdual\in\R$,
\begin{equation}
\left.\langle\varphi_n|\varphi_m\rangle\right|_{\z\mapsto\zdual\in\R}
\equiv(\varphi_n|\varphi_m)=\left(\pi
W(\z')\right)^{m}n!\,\delta_{nm}\;.
\end{equation}

Alternatively, in standard physics notation,
\begin{equation}
\langle \z|n\rangle\equiv\widehat{H}_n(\hat{\z}) =:\langle\z|
H_n|0\rangle
\end{equation}
which introduces the idea of a wave \emph{functional}. And
\begin{equation}
\langle n|m\rangle \equiv\left.\int_{\Z}\langle n|\z\rangle \langle
\z|m\rangle\exp^{-\hat{\z}^2}\DQ{\z}\right|_{\z\mapsto\zdual\in\R}
=\left.\int_{\Z}\widehat{H}_n(\hat{\z})\mathcal{D}\rho_m({\z})
\right|_{\z\mapsto\zdual\in\R}
\end{equation}
where $\langle n|$ and $|n\rangle$ are basis bras and kets in the
associated Hilbert space of states $\mathcal{H}$ (see appendix A).

\subsection{Gauge Potentials}
The same exercise of the preceding subsection must be repeated for gauge
potentials to construct the data
$(\Z_{\mathcal{G}},\Theta_{\mathcal{G}},
Z_{\mathcal{G}},\mathcal{F}_R(\Z_{\mathcal{G}}))$. Of course, the main
issue (and the essence of gauge theory) is to parametrize the moduli
space of gauge potentials. This will not be pursued here since it is
expected to be rather involved.

However, in this regard, it is instructive to briefly investigate the
implications of functionals in the space $\mathcal{F}_R(\Z)$ (recall
(\ref{integral invariant})). Suppose we have a
$\mathcal{D}_{\Theta,Z}\z$ that is invariant under the action of some
Lie group. Then, according to Propsoition \ref{prop. integration by
parts}, integrable functionals in $\mathcal{F}_R(\Z)$ satisfy
\begin{equation}
\int_{\D}\left[\int_{\Z}\frac{\delta F_{\mu}(\z)}{\delta
\z^i(\dd)}\mathcal{D}_{\Theta,Z}\z\right]\Bold{\mathfrak{v}}^i(\dd)
\,d\tau(\dd)=0\;.
\end{equation}
In particular, for Gaussian integrators and translations, this yields
the Schwinger-Dyson equation (ignoring operator ordering issues)
\begin{equation}
\left\langle 1_P\left|\frac{\delta \mathcal{F}\mu(\z)}{\delta \z^i(\dd)}
+\mathcal{F}\mu(\z)\frac{\delta Q(\z)}{\delta
\z^i(\dd)}\right|1_P\right\rangle=0\;.
\end{equation}

Evidently, there exists an equivalence relation among the functionals of
interest. Instead of working with $\mathcal{F}_R(\Z)$, one can try to
construct a nilpotent operator that enforces the equivalence relation
within $\mathcal{F}(\Z)$ and then consider its cohomology. In the proper
geometrical setting, this tack should lead to the BRST formalism within
the CDM scheme.

\subsection{Effective Action}
The Gaussian and Hermite integrators are characterized by quadratic
forms associated with a given action functional. But, it is possible to
characterize an integrator directly in terms of the action functional.
Unfortunately, in most cases, localization of such integrators does not
lead to closed form expressions. In these cases, it is useful to utilize
the functional mean value theorem (Proposition \ref{prop. mean value1}),
which affords alternative calculational methods.

Assume given an action functional $\mathcal{S}(\z)$. Construct an
associated quadratic form $Q$ and define the Gaussian integrator
\begin{equation}
\int_{\Z}\exp^{-2\pi i\zdual}\mathcal{D}\omega(\z) =Z(\z')\;.
\end{equation}
Now characterize a new integrator by the replacement
$Q\rightarrow\mathcal{S}$;
\begin{equation}
\int_{\Z}\exp^{-2\pi i\zdual}\mathcal{D}\widetilde{\omega}(\z)
:=\int_{\Z}\exp^{-2\pi i\zdual}e^{\{\pi
i\mathcal{S}(\z)\}}\mathcal{D}_{\mathcal{S},W_{\mathcal{S}}}(\z)
=:\widetilde{Z}(\z')\;.
\end{equation}
Unless $\mathcal{S}(\z)$ is quadratic, this will localize to complicated
integrals in general.

Apply Proposition \ref{prop. mean value1}: Let
\begin{equation}
\exp{\{\pi i\Gamma(\langle\z \rangle)-2\pi i\langle\z',\langle\z
\rangle\rangle\}}:=\widetilde{Z}(\z')=:\exp{\{-\pi i
W_{\mathcal{S}}(\z')\}}\;.
\end{equation}
and define $\langle\z \rangle:\Z'\rightarrow\Z$ by
\begin{eqnarray}
\langle\z \rangle(\z')(\dd)
&:=&\frac{-\delta}{\delta\z'(\dd)}\ln\mathcal{F}\widetilde{\omega}(\z')\notag\\
&=&\frac{-1}{\widetilde{Z}(\z')}\frac{\delta}{\delta\z'(\dd)} \int_\Z
e^{\{-2\pi
i\langle\z',\z\rangle\}}\mathcal{D}\widetilde{\omega}(\z)\notag\\
&=&-\frac{1}{2}\frac{\delta W_{\mathcal{S}}(\z')}{\delta\z'(\dd)}\;.
\end{eqnarray}
Symmetry of the Fourier transform suggests we also define $\langle\z'
\rangle:\Z\rightarrow\Z'$ by
\begin{eqnarray}
\langle\z' \rangle(\z)(\dd)
&:=&\frac{\delta}{\delta\z(\dd)}\ln\mathcal{F}\widetilde{\mu}(\z)\notag\\
&=&\frac{1}{\widetilde{Y}(\z)}\frac{\delta}{\delta\z(\dd)} \int_{\Z'}
e^{\{2\pi i\langle\z'\z\rangle\}}d\widetilde{\mu}(\z')\notag\\
&=&\frac{1}{2}\frac{\delta \Gamma(\z)}{\delta\z(\dd)}
\end{eqnarray}
where $\widetilde{Y}(\z):=\exp{\{\pi
i\Gamma(\z)\}}:=\mathcal{F}\widetilde{\mu}(\z)$ and
$d\widetilde{\mu}(\z'):=\exp\{-\pi iW_{\mathcal{S}}(\z')\}\,d\mu(\z')$.

In particular,
\begin{equation}
\langle\z(\dd)\rangle_0:=\langle\z\rangle(0)(\dd)/N
=\frac{1}{N}\int_{\Z}\z(\dd)\mathcal{D}\widetilde{\omega}(\z)
=\left.\frac{1}{2}\frac{\delta
W_{\mathcal{S}}(\z')}{\delta\z'(\dd)}\right|_{\z'=0}
\end{equation}
and
\begin{equation}
\langle\z'(\dd)\rangle_0:=\langle\z'\rangle(\langle\z\rangle_0)(\dd)/N
=\frac{1}{N}\int_{\Z'}\z'(\dd)d\widetilde{\mu}(\z')
=\left.\frac{1}{2}\frac{\delta
\Gamma(\z)}{\delta\z(\dd)}\right|_{\z=\langle\z \rangle_0}
\end{equation}
where $N:=\exp{\{-\pi
iW_{\mathcal{S}}(0)\}}=\int_{\Z}\mathcal{D}\widetilde{\omega}(\z)$ is
the normalization of the integrator $\mathcal{D}\widetilde{\omega}(\z)$.
In words, $\langle\z(\dd) \rangle_0$ is the functional average of
$\z(\dd)$ with respect to the integrator
$\mathcal{D}\widetilde{\omega}(\z)$. The normalization factor for
$\langle\z'\rangle_0$ is $N$ since $\exp{\{\pi i\Gamma(\langle\z
\rangle_0)\}}=\exp{\{\pi iW_{\mathcal{S}}(0)\}}$. Clearly,
$\Gamma(\langle\z \rangle)$ is the conventional effective action.

The effective action can be given an interesting characterization in
terms of functionals in $\mathcal{F}_R(\Z)$ and (\ref{symmetric}). Also,
it is fruitful to extend the analysis to the Hermite integrator.
However, these considerations will appear later as they fall outside the
present scope.

\subsection{d+1 Dimensions} In many physics applications
 (canonical quantization in QFT), one is
interested in the time evolution of a dynamical system without gauge
symmetry. In this context, the manifold $\D$ is given a foliation
$\D=\mathbb{X}\times\mathbb{T}$ and some kind of boundary conditions are
specified consistent with a time evolution interpretation. The objects
of interest are
\begin{equation}
\langle F\rangle_{a,b}:=\int_{\F_{a,b}(\M)}F(\f)\mathcal{D}\f
\end{equation}
where $\F_{a,b}(\M)$ is the space of matter fields with initial(final)
boundary condition $\f(\x,\ti_{a(b)})=f_{a(b)}\subset\M$.

Consequently, specialize to the case $\U=\R^d\times\R_+$ and $G$
trivial. Let $\Z_0$ be the space of maps $\z:\U\rightarrow\C^{n+1}$ such
that $\z(\mathrm{z},0)=0$ for all $\mathrm{z}\in\p\U\subset\R^d$. The
parametrization $P:\Z_0\rightarrow\F_{a}(\M)$ is defined point-wise by
the exterior differential system
\begin{equation}
    \left\{ \begin{array}{ll}\Bold{d\f}(\z)(\mathrm{z},\ti)=
    \Xa(\f(\z)(\mathrm{z},\ti))\Bold{d\z}(\mathrm{z},\ti)^{\alpha}\\
    \f(\x,\ti_a)=f_a
    \end{array}
    \right.
\end{equation}
where $f_{a}$ is a fixed submanifold of dimension $d$ in $\M$ and
$\alpha\in\{1,\ldots,n+1\}$. Denote the solution by
$\f(\z)(\mathrm{z},\ti)=f_a\cdot\mathit{\Sigma}(\z(\mathrm{z},\ti))$.
Assume given an action functional $\mathcal{S}(\f)$ and construct the
Hermite integrator. Care must be exercised at this point because
typically the fields have no degrees of freedom in the $\mathbb{T}$
direction for a $d+1$ decomposition: Equivalently, the quadratic form is
degenerate on $\Z_0$. But we require a nondegenerate quadratic form.

Without going into a detailed justification\footnote{Again, a careful
analysis of this situation is outside the scope of this article, and
will be presented elsewhere.}, the `fix' is to consider the space
$\widetilde{\Z}_{0,1}$ of maps
$\widetilde{\z}:\widetilde{\U}=\R^d\times\I\rightarrow\C^n$ such that
$\widetilde{\z}(\mathrm{z},0)=\widetilde{\z}(\mathrm{z},1)=0$ for all
$\mathrm{z}\in\p\widetilde{\U}\subset\R^d$. This space parametrizes
$\F_{a,b}(\M)$ directly. On $\widetilde{\Z}_{0,1}$, the relevant
quadratic form will include the parameter $\mathrm{s}$ that appeared in
section \ref{sec. functional integration}. Recall that $\mathrm{s}$
represents the time interval $[\ti_a,\ti_b]$. (In a very loose sense,
$\widetilde{\Z}_{0}$ can be thought of as arising from a localization
${\Z}_{0}\longrightarrow\widetilde{\Z}_{0}\times\R_+$ induced by the
foliation.)

Finally,
\begin{equation}
\langle F_n\rangle_{a,b}
=\int_{\widetilde{\Z}_{0,1}}F_{\mu}(\widetilde{\z};\mathrm{s})
\mathcal{D}\rho_n(\widetilde{\z})\;,
\end{equation}
where $F_{\mu}(\z;\mathrm{s})\in\mathcal{F}(\widetilde{\Z}_{0,1})$ is
integrable relative to the Hermite volume form.

\section{Summary}
We proposed an extension of the CDM scheme for functional integration
that can be applied to bosonic quantum field theory. There are several
obvious directions for further study. Among them are: including
fermionic fields, a symplectic construction, renormalization,
topologically non-trivial source and target manifolds, implications of
symmetry, and specific applications/examples.

Additionally, this brief study uncovered several aspects that require
further understanding and/or development. At center stage of the
construction is the parametrization linking the space of physical fields
to a relevant Banach space. The facts that path integrals solve partial
differential equations (\cite{LA2}) and the parametrization is specified
by an exterior differential system suggest a possibly fruitful avenue of
investigation. The link between exterior differential systems and the
calculus of variations (\cite{GR}) also deserves investigation. The
de-emphasis of Gaussian integrators suggests the study of new
integrators and their applications. The correspondence between
functional integration and quantum field theory should be expanded: in
particular, the issue of operator ordering needs to be clarified and
further development of the effective action is indicated. Finally, the
whole issue of gauge symmetry and the incorporation of the BRST
framework requires development.

\appendix
\section{Functional Integral/QFT Correspondence}
Quantum field theory supplies (among other things) a complex Hilbert
space $\mathcal{H}$ of states, operators on $\mathcal{H}$, and a basis
of $\mathcal{H}$ that provides a reference for relevant physical
quantities. These are determined to some extent by a choice of an action
functional and (often) the presence of some symmetry.

This appendix exhibits the correspondence between these standard physics
concepts/objects and their functional integral counterparts for the case
when $\Z$ possesses a Hilbert space structure. The exposition is
strictly at a formal level and restricted to $\Theta$ of the form
$\Theta(\z,\z')\sim\theta(\z)\exp\{-2\pi i\zdual\}$. In this case,
integrable functionals under the parametrization $P$ will be of the form
$F_{\mu}(P(\z))\sim
\mathcal{F}\mu(P(\z))\theta(P(\z))=:f(P(\z))\theta(P(\z))$.

The foundation for the correspondence will be the localization
principle: functional integrals over a localized (finite dimensional)
subspace of $\Z$ must reduce to the correct finite dimensional integrals
irrespective of the dimension. This, together with the fact that $\Z$ is
a Hilbert space, suggests the (loose) interpretation of the functional
integral as a kind of `quantum' scalar product denoted
$\langle\cdot|\cdot\rangle$; and, under certain localization conditions,
it should reduce to the standard or `classical' scalar product denoted
$(\cdot|\cdot)$.

Consequently the \emph{functional} scalar product,
\begin{equation}
\langle f_P|g_P\rangle
:=\int_{\Z}f^*(P(\z))g(P(\z))\mathcal{D}\omega(\z)\;.
\end{equation}
where $\mathcal{D}\omega(\z):=\theta(P(\z)) \mathcal{D}_{\Theta,Z}(\z)$,
is interpreted as encoding quantum deformations of the scalar product
$(f_P|g_P)$. The subscript $P$ indicates implicit parametric/boundary
dependence introduced by the parametrization. Accordingly, the subscript
$P$ encodes a particular representation of $\mathcal{H}$.

This `quantum' scalar product clearly reduces to a scalar product
relative to some measure $\omega$ on $\C^n$ under the localization
$\z\mapsto\bigoplus_i^n\langle\z'_i,\z\rangle\in\C^n$;
\begin{eqnarray}
(f_P|g_P)
&:=&\int_{\Z}f^*(P(\zdual))g(P(\zdual))\mathcal{D}\omega(\z)\notag\\
&\rightarrow&\int_{\C^n}f^*_P(\bold{u})g_P(\bold{u})
d\omega(\bold{u})\;.
\end{eqnarray}

In particular, consider the integrable functional associated with the
Dirac measure on $\Z'$, i.e. $f(\z)=1\;\forall \z\in\Z$. Then
\begin{equation}\label{vacuum}
\langle 1_P|1_P\rangle:=\int_{\Z}\mathcal{D}\omega(\z)
\end{equation}
can be interpreted as the `quantum' scalar product associated with
$\z'=0$. In physics terms, $\z'=0$ represents vanishing sources; so it
is reasonable to define $|1_P\rangle$ to be the ground (or vacuum)
state. This allows the familiar identification
\begin{equation}\label{vacuum transition}
\langle 1_P|f_P\rangle \equiv\langle
1_P|f|1_P\rangle:=\int_{\Z}f(P(\z))\mathcal{D}\omega(\z)\;.
\end{equation}

The formal identifications (\ref{vacuum}) and (\ref{vacuum transition})
reduce to their well-known counterparts under localization, and they
imply that the localization defines a representation. Their
interpretation in terms of a `quantum' scalar product suggests the
integrator $\mathcal{D}\omega(\z)$ should be chosen with regard to some
relevant basis. Then the localization principle yields the
correspondence between the (function) inner product on $\mathcal{H}$ and
the (functional) inner product on $\Z$. And, up to operator ordering
issues, (\ref{vacuum transition}) encodes the correspondence between
operators on $\mathcal{H}$ and integrable functionals in
$\mathcal{F}(\Z)$.

\section{Some Integral Properties}
A pervasive theme in the development of functional integration is its
reduction to standard integration theory under localization.  This
appendix lists several important properties of functional integrals in
the CDM scheme that mimic properties of standard integrals.

\begin{proposition}[Linearity]\label{prop. linearity}
The integral operator $\int_{\Z}$ is a linear operator on
$\mathcal{F}(\Z)$.\end{proposition} \emph{Proof.} For $a,b\in\C$;
\begin{eqnarray}\label{linearity}
  \int_{\Z}(aF_{\mu}+bF_{\nu})(\z)\,\mathcal{D}_{\Theta,Z}\z
 &=&\int_{\Z'}Z(\z')\,(ad\mu+bd\nu)(\z')\nonumber\\
 &=&a\int_{\Z'}Z(\z')\,d\mu(\z')+b\int_{X'}Z(\z')\,d\nu(\z')\nonumber\\
 &=&a\int_{\Z}F_{\mu}(\z)\,\mathcal{D}_{\Theta,Z}\z
 +b\int_{\Z}F_{\nu}(\z)\,\mathcal{D}_{\Theta,Z}\z\;.\nonumber\\
\end{eqnarray}
$\QED$

\begin{proposition}[Order of integration]\label{prop. order interchange}
The order of integration over $\Z$ and $\Z'$ can be interchanged.
\end{proposition}
\emph{Proof.} From Definitions \ref{def1}-\ref{def3}, it follows readily
that
\begin{eqnarray}\label{order interchange}
   \int_{\Z}\left[\int_{\Z'}\Theta(\z,\z')\,d\mu(\z')\right]\,\mathcal{D}_{\Theta,Z}\z
   &=&\int_{\Z}F_{\mu}(\z)\,\mathcal{D}_{\Theta,Z}\z\notag\\
   &=&\int_{\Z'}\left[\int_{\Z}\Theta(\z,\z')\,\mathcal{D}_{\Theta,Z}\z\right]\,d\mu(\z')\;.
\end{eqnarray}
$\QED$

\begin{proposition}[Fubini]\label{prop. Fubini}
Suppose that $\Z=\Z_1\otimes\Z_2$. If $\Theta$ and $Z$ are separable,
i.e. $\Theta(\z,\z')=\Theta((\z_1,\z_2),(\z_1',\z_2'))
=\Theta_1(\z_1,\z_1')\cdot\Theta_2(\z_2,\z_2')$ and
$Z(\z')=Z((\z_1',\z_2'))=Z_1(\z_1')\cdot Z_2(\z_2')$, then
\begin{eqnarray}\label{Fubini}
  \int_\Z F_{\mu}(\z)\,\mathcal{D}_{\Theta,Z}\z
  &=&\int_{\Z_1}\left[\int_{\Z_2}F(\z_1,\z_2)
  \,\mathcal{D}_{\Theta_2,Z_2}\z_2\right]\,\mathcal{D}_{\Theta_1,Z_1}\z_1\notag\\
  &=&\int_{\Z_2}\left[\int_{\Z_1}F(\z_1,\z_2)
  \,\mathcal{D}_{\Theta_1,Z_1}\z_1\right]\,\mathcal{D}_{\Theta_2,Z_2}\z_2\;.
\end{eqnarray}\end{proposition}

\emph{Proof.}
\begin{eqnarray}
  \int_\Z F_{\mu}(\z)\,\mathcal{D}_{\Theta,Z}\z
  &=&\int_{\Z'}Z(\z')\,d\mu(\z')\nonumber\\
  &=&\int_{\Z_1'}\int_{\Z_2'}Z_1(\z_1')Z_2(\z_2')\,d\mu(\z_1')\,d\mu(\z_2')
  \nonumber\\
  &=&\int_{\Z_1'}\int_{\Z_2'}\int_{\Z_1}\int_{\Z_2}
  \Theta_1(\z_1,\z_1')\Theta_2(\z_2,\z_2')\nonumber\\
  &&\hspace{1in}\times\,\mathcal{D}_{\Theta_1,Z_1}\z_1
  \,\mathcal{D}_{\Theta_2,Z_2}\z_2\,d\mu(\z_1')\,d\mu(\z_2')\nonumber\\
&=&\int_{\Z'}\int_{\Z_1}\int_{\Z_2}\Theta(\z,\z')
  \,\mathcal{D}_{\Theta_2,Z_2}\z_2\,\mathcal{D}_{\Theta_1,Z_1}\z_1
  \,d\mu(\z')\nonumber\\
&=&\int_{\Z_1}\int_{\Z_2}\int_{\Z'}\Theta(\z,\z')\,d\mu(\z')
  \,\mathcal{D}_{\Theta_2,Z_2}\z_2\,\mathcal{D}_{\Theta_1,Z_1}\z_1\nonumber\\
  &=&\int_{\Z_1}\left[\int_{\Z_2}F(\z_1,\z_2)\,\mathcal{D}_{\Theta_2,Z_2}\z_2\right]
  \,\mathcal{D}_{\Theta_1,Z_1}\z_1
\end{eqnarray}
The fourth equality is a consequence of the Fubini theorem for $\mu$.
The order of integration over $\Z'$ and $\Z$ in the fifth equality can
be interchanged according to Proposition \ref{prop. order interchange}.
$\QED$

\begin{corollary}\label{finite interchange}
For $\Z$ an infinite dimensional separable Banach space and $\M$ a
finite dimensional manifold,
\begin{equation}
  \int_{\Z}\int_{\M}=\int_{\M}\int_{\Z}\;.
\end{equation}
\end{corollary}

\emph{Proof.} Definition \ref{def1} can be used to define a translation
invariant integrator for finite dimensional $\Z=\R^n$ by
\begin{equation}
  \int_{\R^n}\exp^{\{-\pi W_{\R^n}^{-1}(\mathbf{u})-2\pi
i\langle\mathbf{u}',\mathbf{u}\rangle_{\R^n}\}}
|\mathrm{det}\,W_{\R^n}|^{-1/2}\,d\mathbf{u} =\exp^{\{-\pi
W_{\R^n}(\mathbf{u'})\}}
\end{equation}
where $W_{\R^n}$ is a positive definite invertible quadratic form on
$\R^n$. This integrator can be extended to manifolds in the usual
manner. Set $\Z_2=\R^n$ in Proposition \ref{prop. Fubini}; then
$\int_{\Z}\int_{\R^n}=\int_{\R^n}\int_{\Z}$ and, by extension, the
corollary follows.$\QED$

\begin{proposition}[Mean value]\label{prop. mean value1}
Let $\langle \z\rangle := \int_\Z \z\,\mathcal{D}_{\Theta,Z}\z$ and
define $\widetilde{\Theta}(\langle \z\rangle,\cdot):=Z(\cdot)$. Then,
\begin{equation}
  \widetilde{F_{\mu}}(\langle \z\rangle)
  =\int_\Z F_{\mu}(\z)\,\mathcal{D}_{\Theta,Z}\z\;.
\end{equation}
\end{proposition}
\emph{Proof.}
\begin{eqnarray}
  \widetilde{F_{\mu}}(\langle \z\rangle)
  &:=&\int_{\Z'}\widetilde{\Theta}(\langle \z\rangle,\z')d\mu(\z')\nonumber\\
  &=&\int_{\Z'}Z(\z')d\mu(\z')\nonumber\\
  &=&\int_\Z F_{\mu}(\z)\,\mathcal{D}_{\Theta,Z}\z\;.
\end{eqnarray}
$\QED$

\begin{proposition}[Change of variable]\label{prop. change of variable}
Let $M:\mathfrak{X}\rightarrow \mathfrak{Y}$ and
$R:\mathfrak{Y}'\rightarrow \mathfrak{X}'$ be diffeomorphisms. Under the
change of variable $\mathfrak{x}\mapsto M(\mathfrak{x})=\mathfrak{y}$,
\begin{equation}\label{change of variable}
  \int_{\mathfrak{Y}}F_{\mu}(\mathfrak{y})
  \,\mathcal{D}_{\overline{\Theta},\overline{Z}}\mathfrak{y}
   =\int_{\mathfrak{X}}F_{\mu}(M(\mathfrak{x}))
   \,\mathcal{D}_{\Theta,Z}\mathfrak{x}
\end{equation}
where $\overline{\Theta}$ and $\overline{Z}$ are defined in (\ref{theta
bar}) and (\ref{Z bar}).

\end{proposition}
\emph{Proof.} Note that
\begin{eqnarray}\label{F circle M}
  (F_{\mu}\circ M)(\mathfrak{x})
  &=&\int_{\mathfrak{Y}'}\overline{\Theta}(M(\mathfrak{x}),\mathfrak{y}')
  \,d\mu(\mathfrak{y}')\nonumber\\
  &=&\int_{\mathfrak{X}'}\overline{\Theta}
  (M(\mathfrak{x}),R^{-1}(\mathfrak{x}'))
  \,d\mu(R^{-1}(\mathfrak{x}'))\nonumber\\
  &=&\int_{\mathfrak{X}'}\overline{\Theta}
  (M(\mathfrak{x}),R^{-1}(\mathfrak{x}'))
  \,d\nu(\mathfrak{x}')\nonumber\\
  &=&\int_{\mathfrak{X}'}\Theta(\mathfrak{x},\mathfrak{x}')
  \,d\nu(\mathfrak{x}')\nonumber\\
  &=&F_{\nu}(\mathfrak{x})\;.
\end{eqnarray}
Hence,
\begin{eqnarray}
  \int_{\mathfrak{Y}}F_{\mu}(\mathfrak{y})
  \,\mathcal{D}_{\overline{\Theta},\overline{Z}}\mathfrak{y}
   &=&\int_{\mathfrak{Y}'}\overline{Z}(\mathfrak{y}')\,d\mu(\mathfrak{y}')\nonumber\\
  &=&\int_{\mathfrak{X}'}(\overline{Z}\circ R^{-1})(\mathfrak{x}')
  \,d\mu(R^{-1}(\mathfrak{x}'))\nonumber\\
  &=&\int_{\mathfrak{X}'}Z(\mathfrak{x}')
  \,d\nu(\z')\nonumber\\
   &=&\int_{\mathfrak{X}}F_{\nu}(\mathfrak{x})
   \,\mathcal{D}_{\Theta,Z}\mathfrak{x}\nonumber\\
   &=&\int_{\mathfrak{X}}F_{\mu}(M(\mathfrak{x}))
   \,\mathcal{D}_{\Theta,Z}\mathfrak{x}\;.
\end{eqnarray}$\QED$

\begin{corollary}\label{cor. change of variable}
If $M(\Z)=\Z$ and $M'$ is nuclear and $F_{\mu}(\z)\in\mathcal{F}_R(\Z)$,
then
\begin{equation}\label{Invariant integrator}
\int_{\Z}F_{\mu}(\widetilde{\z})
\,\mathcal{D}_{\overline{\Theta},\overline{Z}}\widetilde{\z}
   =\int_{\Z}F_{\mu}(M(\z))(\mathrm{Det}M'_{(\z)})
   \,\mathcal{D}_{\overline{\Theta},\overline{Z}}\z\;.
\end{equation}
\end{corollary}

\emph{Proof.} By (\ref{theta relation eq.}),
\begin{eqnarray}
   \int_{M(\Z)}F_{\mu}(\widetilde{\z})
   \,\mathcal{D}_{\overline{\Theta},\overline{Z}}\widetilde{\z}
   &=&\int_{R^{-1}(\Z')}\overline{Z}(\widetilde{\z}')\,d\mu(\widetilde{\z}')\notag\\
  &=&\int_{\Z'}Z(\z')\,d\nu(\z')\notag\\
  &=&\int_{\Z'}(\mathrm{Det}R'_{(R^{-1}(\z'))})\overline{Z}(\z')\,d\nu(\z')\notag\\
  &=&\int_\Z F_{\nu}(\z)(\mathrm{Det}M'_{(\z)})
  \,\mathcal{D}_{\overline{\Theta},\overline{Z}} \z\notag\\
  &=&\int_\Z F_{\mu}(M(\z))(\mathrm{Det}M'_{(\z)})
  \,\mathcal{D}_{\overline{\Theta},\overline{Z}} \z\;.
\end{eqnarray}
The fourth line follows from the third by Definitions
\ref{def1}--\ref{def3}. $\QED$

\begin{proposition}[Integration by parts]\label{prop. integration by parts}
Suppose that $F_{\mu}\in \mathcal{F}_R^{\wedge}(\Z)$, then
\begin{equation}
\int_{\Z}\mathrm{i}_{\boldmath{\mathfrak{v}}} d[F_{\mu}(\z)]
\;\mathcal{D}_{\Theta,Z}\z
 =-\int_{\Z}F_{\mu}(\z)\,d[\mathrm{i}_{\boldmath{\mathfrak{v}}}
 \mathcal{D}_{\Theta,Z}\z]\;.
\end{equation}
In particular, if
$\mathcal{L}_{\boldmath{\mathfrak{v}}}\mathcal{D}_{\Theta,Z}\z=0$, then
\begin{equation}\label{integration by parts}
0=\int_{\D}\left[\int_{\Z}\frac{\delta F_{\mu}(\z)}{\delta \z(\dd)}
\;\mathcal{D}_{\Theta,Z}\z\right]\Bold{\mathfrak{v}}^i(\dd)\,d\tau(\dd)
\;.
\end{equation}
\end{proposition}

\emph{Proof.} The first equality follows trivially from Restriction
\ref{top form} and $\mathcal{L}_{\boldmath{\mathfrak{v}}}
=d\mathrm{i}_{\boldmath{\mathfrak{v}}}+\mathrm{i}_{\boldmath{\mathfrak{v}}}d$.
The second expression follows from Proposition \ref{prop. Fubini} and
\begin{equation}
\mathrm{i}_{\boldmath{\mathfrak{v}}}\Bold{dF}_{\mu}(\z)
=\lim_{\xi\rightarrow 0}\frac{F_{\mu}(\sigma_{g(\xi)}\z)
-F_{\mu}(\z)}{\xi} =\int_{\D}\frac{\delta F_{\mu}(\z)}{\delta
\z^i(\dd)}\Bold{\mathfrak{v}}^i(\dd)\,d\tau(\dd)
\end{equation}
where
\begin{equation}
\Bold{\mathfrak{v}}=\lim_{\xi\rightarrow
0}\frac{\sigma_{g(\xi)}\z-\z}{\xi}\;.
\end{equation} $\QED$


\end{document}